\documentclass[aps, prd, onecolumn, tightenlines, notitlepage, superscriptaddress, nofootinbib, preprintnumbers, floatfix,showkeys,11pt,altaffilletter]{revtex4-1}

\usepackage[official]{eurosym}
\usepackage[normalem]{ulem}
\usepackage{amstext}
\usepackage{graphicx}
\usepackage{url}
\usepackage{color}
\usepackage{ulem}
\usepackage[version=4]{mhchem}
\usepackage[utf8]{inputenc}
\usepackage{fontawesome}
\usepackage{yfonts}

\pdfoutput=1
\usepackage{textcomp}
\usepackage{comment}
\usepackage{yfonts}
\usepackage{epsfig,amsfonts,mathrsfs,amsmath,amssymb,graphicx,color,slashed,multirow}
\usepackage{amsmath,latexsym,amssymb,graphicx,slashed,color,enumerate,url,cancel,gensymb}
\usepackage{textcomp}

\usepackage[x11names]{xcolor}
\usepackage[colorlinks,pdfstartview=FitV,breaklinks=true]{hyperref}

\usepackage{booktabs}
\usepackage{adjustbox}
\usepackage{textgreek} 
\usepackage{booktabs} 
\usepackage{float}

\usepackage{lmodern}
\usepackage{ae,aecompl}
\usepackage{appendix}


\newcommand{\eq}[1]{(\ref{#1})}

\definecolor{vdrgreen}{rgb}{0.0, 0.7, 0.0}

\definecolor{lightapricot}{rgb}{0.99, 0.84, 0.69}
\definecolor{nicered}{rgb}{0.7,0.1,0.1}
\definecolor{nicegreen}{rgb}{0.1,0.5,0.1}
\definecolor{coral}{rgb}{1.0, 0.5, 0.31}
\definecolor{blue(ncs)}{rgb}{0.0, 0.53, 0.74}
\definecolor{darkspringgreen}{rgb}{0.09, 0.45, 0.27}
\definecolor{seagreen}{rgb}{0.18, 0.55, 0.34}

\definecolor{cadmiumgreen}{rgb}{0.0, 0.42, 0.24}
\definecolor{chromeyellow}{rgb}{1.0, 0.65, 0.0}
\definecolor{darkturquoise}{rgb}{0.0, 0.81, 0.82}
\definecolor{denim}{rgb}{0.08, 0.38, 0.74}
\definecolor{purple(x11)}{rgb}{0.63, 0.36, 0.94}
\definecolor{red(ncs)}{rgb}{0.77, 0.01, 0.2}
\definecolor{ruddypink}{rgb}{0.88, 0.56, 0.59}
\definecolor{slateblue}{rgb}{0.42, 0.35, 0.8}
\definecolor{airforceblue}{rgb}{0.36, 0.54, 0.66}
\definecolor{orange(colorwheel)}{rgb}{1.0, 0.5, 0.0}

\def\mp{M_{\rm P}} 
\def\aap{\ref@jnl{A\&A}}                

\makeatletter
    \newcommand{\colorboxed}[3][white]{\fcolorbox{#2}{#1}{\m@th$\displaystyle#3$}}
\makeatother

\AtBeginDocument{\hypersetup{citecolor=black,linkcolor=black,urlcolor=black}}
\usepackage{appendix}

\def \aap  {A\&A}

\begin{document}

\count\footins = 1000

\preprint{IFT-UAM/CSIC-24-84\quad\quad\quad\quad\quad\quad\quad\quad\quad\quad\quad\quad\quad\quad\quad\quad\quad\quad\quad\quad\quad\quad\quad \quad\quad\quad\quad\quad  LAPTH-031/24}
\title{{\LARGE {Primordial black hole formation from\\ \vspace{0.17cm} self-resonant preheating?}}}

\author{Guillermo Ballesteros}
\email{guillermo.ballesteros@uam.es}
\affiliation{Departamento de F\'isica T\'eorica, Universidad Aut\'onoma de Madrid (UAM),
Campus de Cantoblanco, 28049 Madrid, Spain}
\affiliation{Instituto de F\'isica Te\'orica UAM-CSIC, Campus de Cantoblanco, 28049 Madrid, Spain}

\author{Joaquim Iguaz Juan}
\email{jiguazjuan@umass.edu}
\affiliation{Amherst Center for Fundamental Interactions, Department of Physics, University of Massachusetts,
1126 Lederle Graduate Research Tower, Amherst, MA 01003-9337, U.S.A.}
\affiliation{LAPTh, CNRS and Univ. of Savoie Mont Blanc, 9 chemin de Bellevue - BP 110, 74941 Annecy-Le-Vieux, France}

\author{Pasquale D. Serpico}
\email{pasquale.serpico@lapth.cnrs.fr}
\affiliation{LAPTh, CNRS and Univ. of Savoie Mont Blanc, 9 chemin de Bellevue - BP 110, 74941 Annecy-Le-Vieux, France}

\author{Marco Taoso}
\email{marco.taoso@to.infn.it}
\affiliation{Istituto Nazionale di Fisica Nucleare (INFN), Sezione di Torino, Via P. Giuria 1, I--10125 Torino, Italy}

\begin{abstract}
We revisit the question of how generic is the formation of primordial black holes via self-resonant growth of inflaton fluctuations in the post-inflationary, preheating phase. 
Using analytical and lattice calculations,  
we find that primordial black hole production is far from being a generic outcome. Also, in most of the parameter space of viable inflationary models, the metric preheating term is subleading to the anharmonic terms and the approximation of a quadratic potential for describing the resonance dynamics is inadequate. Nonetheless, the anharmonicity of the potential cannot be used to rescue the mechanism: The generic outcome of the non-linear evolution of the scalar field in this case is the formation of metastable transients or oscillons, that do not generically collapse into black holes.   
\end{abstract}

\maketitle

\tableofcontents

\section{Introduction}

Different mechanisms have been proposed for primordial black hole (PBH)  production in the early Universe. The most popular one is the gravitational collapse of large primordial overdensities generated in non-standard inflationary scenarios, drastically departing from standard slow-roll inflation. In this context, a fluctuation with a density contrast larger than a critical threshold $\delta_c \lesssim\mathcal{O}(1)$ collapses soon after the main Fourier mode that describes it re-enters the particle horizon, leading to a PBH whose mass is proportional to the mass enclosed within the Hubble radius at that time. 
Therefore, in order to produce a substantial amount of PBHs, the variance of scalar fluctuations at small cosmological scales has to be enhanced by several orders of magnitude with respect to the one at the large scales probed by Cosmic Microwave Background (CMB) observations.
Examples of models of inflation overcoming this difficulty, at the expense of fine-tuned parameters, can be found in 
reviews such as~\cite{Sasaki:2018dmp,Escriva:2022duf,Ozsoy:2023ryl}.   
Alternatively, PBH formation could only involve scales that are always sub-Hubble, 
for instance via bubble collisions in first order phase transitions  or the dynamics of topological defects formed in association to a symmetry-breaking pattern in the early universe,
as speculated already in the Eighties~\cite{Hawking:1982ga,Hawking:1987bn}. These scenarios, which do not necessarily avoid fine-tuning, require anyway beyond-the-standard model ingredients that make them non-generic.
It has also been suggested that PBHs could be formed during preheating, the initial, non-perturbative phase of reheating which follows the inflationary era (see~\cite{Lozanov:2019jxc, Amin:2014eta, Allahverdi:2010xz} for reviews). 
During preheating the perturbations of the inflation field (and/or any field it couples to) can be greatly enhanced, and it has been claimed that this dynamics can lead to the generation of a large population of PBHs~\cite{Bassett:1998wg,Green:2000he,Bassett:2000ha,Suyama:2004mz,Suyama:2006sr,Jedamzik:2010dq}. 

More recently, the matter has received renewed attention, specifically in the context of so-called metric preheating in single-field inflation. Recent works have argued that PBH formation from metric preheating may be possible \cite{Auclair:2020csm} or even unavoidable \cite{Martin:2019nuw} under mild and rather generic conditions. This would be in stark contrast with the tuning that is generically required to produce them in models in which they originate from large curvature fluctuations generated {\it during} inflation.
 Moreover, these articles conclude that the mechanism is so efficient that those PBHs can dominate the energy density of the Universe after inflation and reheating can proceed thanks to Hawking evaporation. 
The PBH originating from preheating would have masses too small to account for the current dark matter --
a strong motivation for current PBH studies, see \cite{Carr:2020xqk,Green:2020jor,Sasaki:2018dmp,Escriva:2022duf,Villanueva-Domingo:2021spv} for recent reviews including bounds on the abundance of these objects~--, as they would have evaporated before today. Nevertheless, their interest for cosmology remains, 
e.g.\ they could alter the evolution of the Universe with respect to the standard picture and even be catalysers for the generation of the {dark matter \cite{Khlopov:2004tn, Fujita:2014hha, Lennon:2017tqq, Hooper:2019gtx, Baldes:2020nuv, Bernal:2021bbv, Masina:2021zpu}.}
In particular, they would suggest the possibility of a much more important role for PBHs than it is currently accepted.

In this work, we first revisit these claims that  metric preheating leads to generic PBH production thanks to the enhanced growth of  scalar fluctuations. Then we address the question of whether a generalised scenario for self-resonant formation of PBH during reheating is viable. Our study implies that PBH formation from preheating dynamics is challenging and certainly non-generic. For busy readers that are familiar with the topic, we find it convenient to summarize now briefly the essence of the logic that we follow to arrive to our conclusion. We first show that the mechanism discussed in \cite{Jedamzik:2010dq,Martin:2019nuw,Auclair:2020csm} purportedly leading to PBH formation is nothing but the linear growth of subhorizon matter fluctuations during an epoch of matter domination. Such fluctuations grow in time proportional to the scale factor of the Universe, while the curvature fluctuations remain constant. In order to draw a reliable conclusion about PBH formation in such a situation, the late-time extrapolation of the linear result is insufficient, and the existing treatments resorting to too idealized conditions, such as spherical symmetry and absence of feedback, are unreliable. Yet, the linear conservation of the curvature fluctuation already indicates that the gravitational potential is unlikely to reach the necessary values for collapse into PBHs, unless this outcome is generically rescued by non-linear effects. A shortcoming of the previous studies~\cite{Martin:2019nuw,Auclair:2020csm} is that a purely quadratic potential was considered. Such a potential is strictly speaking incompatible with the CMB, hence it can only be considered as a limiting case embedded in a more complex potential. It turns out that the realistic viable choices of the potential are relevant to describe the last stages of the evolution of the inflaton and might also rescue the mechanism we are interested in, since (as we show) anharmonic effects are important in its dynamics. One may hope that these effects may enhance the growth of fluctuations during preheating, possibly favoring PBH formation, but it turns out that this is not the case. We see this by doing (non-linear) lattice computations of the evolution of the fluctutions for potentials that are known to provide a good fit to the CMB. In agreement with existing literature on the formation of non-perturbative objects in the post-inflationary epoch, the growth of the inflaton perturbations is quenched by their own backreaction, making PBH formation non-generic at best. Only the formation of fuzzier objects ({\it oscillons}) may be expected.  A final hope for PBHs may be searched within the context of an analysis with full General Relativity, that would extend our lattice study. However, studies in that direction have found that PBHs could only be hoped to form under very particular initial conditions \cite{Aurrekoetxea:2023jwd}, very different from the envisaged growth of generic small inflationary seeds subject to self-resonant enhancement.

Our analysis (following the logic just described) is structured as follows.
In Section~\ref{sec:resonance} we review the mechanism of parametric resonance during preheating and discuss the evolution of the inflation perturbations in a toy model based on a purely quadratic monomial inflationary potential, as it was considered in \cite{Martin:2019nuw,Auclair:2020csm}. In Sec.~\ref{issues} we discuss the observational limitations of this scenario as well as its inability to produce PBHs, in the light of existing literature. 
In Section~\ref{sec:inflamodel} we focus on  more realistic scenarios: On the one hand, we show that accounting for anharmonic terms is typically needed (and these terms are dominant with respect to gravitational ones) in a vast part of the parameter space of viable inflationary models, including popular benchmark ones like Starobinsky inflation \cite{Starobinsky:1980te}.  Thus, the simple scenario described in Sec.~\ref{sec:resonance} is not only strictly speaking ruled out, but cannot be considered as representative of inflationary models, resulting far from generic. On the other hand,  one might wonder if  anharmonic terms may rescue PBH production from preheating in a modified version of the scenario. 
Specifically, one may ask if inflaton self-interactions may boost the formation of non-linear structures if compared to pure gravitational interaction, thus easing PBH formation. 
After studying the dynamics in the linear regime (Sec.~\ref{sec:inflamodel}), in Section~\ref{sec:lattice} we follow the evolution of the perturbations in the nonlinear regime with the help of lattice simulations. As we already mentioned, we find that the parameter space where the formation of non-linear structures is significantly enhanced corresponds to the one where these objects are metastable {\it transients or oscillons} (see e.g.{~\cite{Bogolyubsky:1976yu,Gleiser:1993pt,Copeland:1995fq,Honda:2001xg,Saffin:2006yk,Hindmarsh:2007jb,Amin:2010jq,Amin:2010dc}), rather than PBHs: in this range, we recover results previously obtained in the context of early universe oscillon physics. 
In Section~\ref{sec:conclusions}, we review our results and present our conclusions. Some considerations on the quantitative reliability with which the post-inflationary expansion can be approximated as matter-like are reported in the appendix.


\section{Parametric resonance for the quadratic potential: Metric preheating}
\label{sec:resonance}
We assume inflation to be driven by a single scalar field, $\phi$, with a canonical kinetic term, a potential $V$ and minimally coupled to gravity. The homogeneous background dynamics during and after inflation is governed by the following two equations
\begin{eqnarray}
H^2&=&\frac{1}{3\mp^2}\left(\frac{\dot \phi^2}{2}+V\right)\label{Fried1}\,,\\
\dot H&=&-\frac{\dot\phi^2}{2\mp^2}\,,\label{Fried2}
\end{eqnarray}
where $\mp=(8\pi G)^{-1/2}$ is the reduced Planck mass, $H=\dot{a}/a$ is the Hubble rate of expansion, $a$ is the scale factor of the Universe and dots denote derivatives with respect to cosmic time, $t$. In order to describe the background evolution, but also to understand the dynamics of small fluctuations about it, it is useful to define a sequence of slow-roll functions ({\it parameters}) of $t$. We start defining  $\epsilon_0\equiv1/H$ and then define 
$H\,\epsilon_{i+1} = \dot\epsilon_i/\epsilon_i$, with, in particular, $\epsilon\equiv\epsilon_1= -\dot H/H^2$. If $|\epsilon_{1,2}|\ll 1$, it is easy to see that eq.\ \eq{Fried1} becomes, approximately, $3 \mp^2 H^2-V\simeq 0 $ and the {(first) potential slow-roll parameter} $\epsilon_V\equiv \mp^2( V_{\rm,\phi}/V)^2/2$ satisfies $\epsilon_V \simeq \epsilon$. This is the slow-roll regime when, in addition, $3H\dot\phi+{\rm d}V/{\rm d}\phi\simeq 0$.  The number of e-folds of expansion is defined as ${\rm d}N\equiv H{\rm d}t$, so that ${\rm d}N= {\rm d}\phi/(\sqrt{2\epsilon}\mp)$.
The end of inflation corresponds to the condition $\ddot a=0\Leftrightarrow\epsilon=1$, and we denote the number of e-folds elapsed between some arbitrary reference time\footnote{More precisely, for the following quantitative results we set this time to be the one at which the Fourier mode of the fluctuations with comoving wavenumber equal to $0.05$ Mpc$^{-1}$ becomes super-Hubble during inflation.} and the end of inflation by $N_*$.

We are particularly interested in the dynamics of the (single) inflaton field after inflation, when the field homogeneous  background oscillates around the minimum of its potential. As we describe below, these oscillations induce a growth of the density fluctuations on {\it sub-horizon} scales. In a number of articles, this process (or a related-one) has been argued to enable the formation of PBHs,  see \cite{Jedamzik:2010dq,Martin:2019nuw,Auclair:2020csm,del-Corral:2023apl}. The well-known Mukhanov-Sasaki equation describing the time evolution of the comoving curvature perturbation $\mathcal{R}$  in Fourier space in linear perturbation theory is usually written as\footnote{In the Newtonian gauge, d$s^2 = (1+2\Psi)$d$t^2-a^2(t)(1-2\Phi)$d$\mathbf{x}^2$, 
$\mathcal{R}=-\Phi-H\delta \phi/\dot \phi$.
}
\begin{align} \label{MS}
v''+\left(k^2-\frac{z''}{z}\right)v=0\,,
\end{align}
where  $k$ is the comoving wavenumber, primes are derivatives with respect to conformal time $
\tau$ ($a\, {\rm d}\tau = {\rm d}t$) and 
\begin{align}\label{vDef}
v=z\,\mathcal{R}=\sqrt{2\epsilon}\,\mp\, a\, \mathcal{R}\,.
\end{align}
This equation shows that, in the adequate variables, the dynamics of the linear scalar fluctuations behave as an harmonic oscillator with a time- and scale-dependent frequency. In terms of the slow-roll parameters we defined above, the ratio $z''/z$ reads
\begin{align}\label{zepsilons}
\frac{z''}{z}=a^2H^2\left( 2-\left(1+\frac{\epsilon_2}{2}\right)\epsilon+\left(3+\frac{\epsilon_2}{2}+\epsilon_{3}\right)\frac{\epsilon_2}{2}\right)\,.
\end{align}
 For our purposes, it is convenient to define the variable 
\begin{align}\label{vtilde}
\tilde v \equiv \sqrt{a}\, v = \sqrt{2\epsilon}\,\mp\, a^{3/2}\, \mathcal{R}\,,
\end{align} 
in terms of which, eq.\ \eq{MS} can be recast into the following form, see e.g.\ \cite{Finelli:1998bu}
\begin{align}\label{vtildeMS}
\ddot{\tilde v} +\omega^2(k)\tilde v=0\,,
\end{align}
where
\begin{align}\label{omegavtilde}
\omega^2\equiv\frac{k^2}{a^2} -H^2\left[ \frac{9}{4}+\frac{\epsilon_2}{2}\left(3+\frac{\epsilon_2}{2}+\epsilon_3\right)-\frac{1}{2}\epsilon\left(3+\epsilon_2\right)\right]\,,\quad
\end{align}
or, explicitly in terms of the potential $V$ and $\dot \phi$, as
\begin{align} 
\omega^2
= \frac{{\rm d}^2V}{{\rm d}\phi^2} + \frac{k^{2}}{a^{2}} + \frac{2}{M_{\rm P}^{2}} \frac{{\rm d}V}{{\rm d}\phi}\frac{ \dot{\phi}}{H} + \frac{3 \dot{\phi}^{2}}{M_{\rm P}^{2}} - \frac{\dot{\phi}^{4}}{2 H^{2} M_{\rm P}^{4}} + \frac{3}{4 M_{\rm P}^{2}} P\,,\label{Qfull1}
\end{align}
where $P=\dot{\phi}^2/2-V$ denotes the background pressure and we stress that the last three terms in (\ref{Qfull1}) originate from the fluctuations of the metric.

A quadratic potential, $V=\frac{1}{2}m^2\phi^2$, is often used to approximate the dynamics of the inflaton background during reheating. This approximation can be useful provided that the oscillations of the inflaton condensate about the minimum of the potential, once inflation has ended, are small enough. Assuming such a potential, the condition $\epsilon=1$ gives the value of $H$ at the end of inflation, $H_{\rm end}=m\,\phi_{\rm end}/(2 \mp)$. Then, approximating $\epsilon$ by $\epsilon_V$, we get $\phi_{\rm end}\simeq\sqrt{2}\mp$, which leads to $H_{\rm end}\simeq m/{\sqrt{2}}$. If we denote with $t_{\rm end}$  the (cosmic) time at which inflation ends, 
the following {\it asymptotic} solution of the homogeneous inflaton field equation $\ddot \phi+3H\dot \phi +{\rm d}V/{\rm d}\phi=0$ holds for times $t\gg t_{\rm end}$ so that $H\ll m$:
\begin{equation}
\phi(t)=\phi_{\rm end}\left(\frac{a_{\rm end}}{a}\right)^{3/2}\cos [m\,(t-t_{\rm end})]\label{as_solution}\,.
\end{equation}
One can then compute explicitly the energy density and pressure of the field, organising the terms in powers of $H/m$. Neglecting the slow evolution of $H$ with respect the rapid oscillations with frequency $m$ (WKB approximation), the scale factor evolves as if the Universe was filled with a pressureless matter fluid, as known since the Eighties~\cite{Turner:1983he} (see Appendix~\ref{formulae}).

 \begin{figure}[t]
\centering
\includegraphics[width=0.75\textwidth]{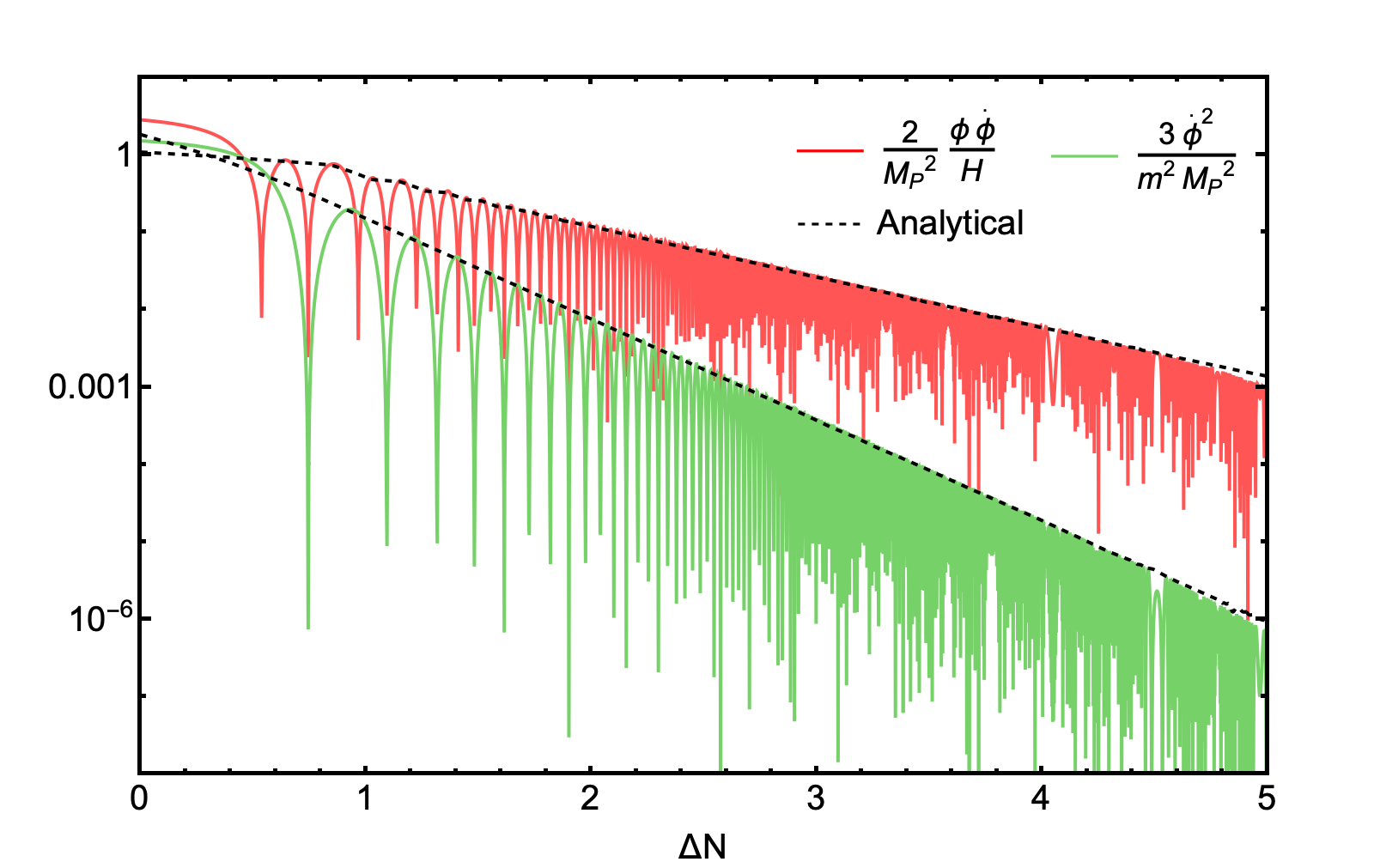} 
\caption{
Post-inflationary time evolution of some representative terms in the parenthesis in Eq.~\eqref{Qfull1}, normalised to $m^2$, for a quadratic potential with $m=1.56 \times 10^{13}$GeV. 
We have imposed $N_{*}=55$ and a normalization of the power spectrum of curvature perturbations as explained in Sec.\ref{sec:Tmodel}.
The numerical results are reported in red and green for the third and fourth term in Eq.~\eqref{Qfull1}, respectively. The dashed lines indicate the analytical result obtained with eq.~\eqref{as_solution}, where we numerically solved for the scale factor.}
\label{termsQuadPot}
\end{figure}

Within the regime of validity of the approximate solution eq.~\eqref{as_solution}, the last three terms on the RHS of eq.\ \eq{Qfull1} redshift as $a^{-3}$ and can be neglected at sufficiently late times because the third one scales as $a^{-3/2}$, {see also \cite{Jedamzik:2010dq}}. In Figure~\ref{termsQuadPot} we verify this statement numerically  for a reasonable choice of parameters by comparing the third term (red) with the fourth one (green), normalised to $m^2$ (second derivative of the potential). The colored lines show the full numerical solutions, while the dashed lines  have been obtained by solving numerically the evolution of the scale factor and plugging this quantity in the analytical approximation of eq.~\eqref{as_solution}.
A few e-folds past the time at which $\epsilon=1$ (when $\Delta N=0$) are shown; this roughly corresponds to $\phi$ evolving from $\sim \mp$ to $\sim 10^{-4}\,\mp$.

  The third term on the RHS of eq.\ \eq{Qfull1} is the one responsible for the {\it metric} instability growth at small cosmological scales. Inserting the solution of eq.~\eqref{as_solution} into eq.s~\eqref{vtildeMS}, \eqref{omegavtilde}, and performing the change of variables $s\equiv m (t-t_{\rm end}) + 3\pi/4$, neglecting all terms other than the third one, the following equation is obtained:
\begin{equation}
\frac{\text{d}^{2}\tilde{v}}{\text{d}s^{2}} + \bigg[ A(k) - 2\, q\, \text{cos}(2\, s) \bigg] \tilde{v} = 0\,.
\label{quadmathieu}
\end{equation}
Assuming that  $a$ is constant over a time $1/m$, this equation is of Mathieu's kind, with ({see also \cite{Jedamzik:2010dq}})
\begin{equation}
A(k) = 1 + \frac{k^{2}}{m^{2} a^{2}} \quad \text{and} \quad q = \frac{\phi_{\rm end}}{M_{\rm P}} \left( \frac{a_{\rm end}}{a} \right)^{3/2}\,.
\label{Aq}\end{equation}
Here we have assumed that the equation of state parameter vanishes immediately after $t_{\rm end}$, which is an approximation leading to ${\cal O}$(1) errors for $t\sim t_{\rm end}$, much like eq.~\eqref{as_solution}.  While this leads to a continuous field value $\phi$ at time $t_{\rm end}$, an alternative convention with comparable errors would be to assume continuity of $H$ (and discontinuity of $\phi$) at $t_{\rm end}$.

 For $A(k)$ and $q$ constant, Eq.~\eqref{quadmathieu} is familiar in inflationary preheating theory~\cite{Kofman:1994rk,Kofman:1997yn}; it admits solutions of the form~\cite{Amin:2014eta,lozanov2019lectures} 
\begin{equation}
\tilde{v}_k = e^{\mu_{k} s} \mathcal{P}_{k+}(s) + e^{-\mu_{k} s} \mathcal{P}_{k-}(s),
\label{solmathieu}
\end{equation}
with $\mathcal{P}_{k\pm}$ periodic functions of $s$ with period equal to the one of the oscillating condensate, $T=2\pi/m$; $\mu_{k}$ is the complex-valued Floquet exponent and depends on both $A$ and $q$. Regions of the parameter space $\{A, q\}$ where  $\Re(\mu_k) \neq 0$, usually  referred to as {\it bands}, correspond to an exponential increase with time of one of the solutions. This phenomenon is dubbed parametric resonance and describes  a fast growth of perturbations with time at the expense of the inflaton condensate oscillating at the bottom of the potential.

When $q\gtrsim 1$, one is in the {\it broad resonance} regime, with $\Re(\mu_k)\sim (2\pi)^{-1}$; the inflaton quanta are produced explosively, in bursts, during only a small fraction of each oscillation cycle of the condensate. Within just a few oscillations, the condensate then disappears~\cite{Kofman:1994rk,Kofman:1997yn}.

When inflation ends with $\phi_{\rm end}\ll \mp$ and in any case at late times,  one has $|q|\ll 1$, hence one can apply Floquet's theory in the {\it narrow} resonance regime~\cite{lozanov2019lectures}. In this case, the first and most important  instability band is given by $1-|q| < A(k) <1+|q|$, singling out
\begin{equation}
\frac{k}{a} < \sqrt{3 H m}\,
\label{condself1}
\end{equation}
as the first region in $k$-space (i.e.\ the largest spatial scales) where the instability takes place.
Due to the expansion of the Universe,  $A$ and $q$ are time-dependent parameters. Strictly speaking, Floquet theory is only applicable when $a$ is constant. In that case, the instability truly represents an exponentially fast growth. Once the time dependence of $a$ is accounted for, apart for the only approximate validity of eq.~\eqref{quadmathieu}, two major consequences are that physical wavenumbers $k/a(t)$ describe trajectories in the Floquet charts,  redshifting into the low-wavenumber region as time goes by; further, the Hubble expansion damps the growth, which is fast only when $\Re(\mu_{k})m/H\gg 1$~\cite{Amin:2014eta,lozanov2019lectures}. 
In the case of interest, on sub-Hubble scales, for modes satisfying
\begin{equation}
H<\frac{k}{a} < \sqrt{3 H m}\,,
\label{sub-hor-inst}
\end{equation}
the narrow resonance condition\footnote{In fact, the narrow resonance condition $q\ll 1$ is not fulfilled soon after the end of inflation, when $q_{\rm end}\simeq \sqrt{2}$. The reliability of the adiabatic approximation in the linear theory around the unaffected background homogeneous field is then unclear, being controlled by the time- and mode-varying value of $\Re(\mu_{k})m/H$. The applicability of the analytical theory assuming $q\ll 1$ everywhere in articles such as~\cite{Jedamzik:2010dq} thus relies fully on the numerical solution.} yields $\Re(\mu_k)=|q|/2$; the growth of $\tilde{v}_k$ can 
be heuristically obtained by the ``adiabatic'' approximation $\tilde{v}_k\propto \exp({\rm d} s \int \mu_k)\propto a^{3/2}$~\cite{Finelli:1998bu,Jedamzik:2010dq,Hertzberg:2014jza},
hence one derives that the density contrast grows as $a$ in the perturbative regime~\cite{Nambu:1996gf,Finelli:1998bu,Jedamzik:2010dq}, recovering the behaviour of dust in Newtonian linear perturbation theory, as it was remarked almost forty years  ago~\cite{1985MNRAS.215..575K}. 

The link between the density perturbation in the Newtonian gauge and the curvature perturbation can be used to deduce that density as well as  $\mathcal{R}$ stay constant on super-Hubble scales (see e.g. \cite{Martin:2020fgl}).  The curvature perturbation $\mathcal{R}$ also stays constant on sub-Hubble scales, at the linear level. Note that the  {\it Jeans wavenumber} $k_J$ in an expanding Newtonian fluid can be computed as
\begin{equation}
\frac{k_J}{a}\equiv \sqrt{\frac{4\pi G_N {\rho}}{c_s^2}}=\sqrt{\frac{3}{2}}\frac{H}{c_s}
\simeq \sqrt{\frac{2}{3}}m\,,\label{Jeans}
\end{equation}
where $c_s$ is the speed of sound (see Appendix~\ref{formulae} for its definition in this case).
Since we assumed $H\ll m$, the modes subject to the condition of eq.~\eqref{sub-hor-inst} are a subset of those allowed by the Jeans analysis ($k\ll k_J$).

\subsection{Is the metric preheating instability leading to PBH formation?}\label{issues}
There exists consensus in the literature about the previous results, 
at least within the regime of validity of the approximations made (more on their realism/plausibility later). 
However, at times it has been the basis for much bolder statements concerning the fate of these growing density contrast modes, which have been  argued to lead  quite generically to PBH formation.\footnote{In order to avoid confusion, let us clarify that the mechanism of interest here is not the same considered e.g. in~\cite{Green:2000he,Bassett:2000ha}, which instead concerns two-scalar field models, and the modes considered for collapse are Hubble-sized. Lattice studies have shown, by the way, that even in this case PBH production is typically not relevant~\cite{Suyama:2004mz}; see however~\cite{Torres-Lomas:2014bua} for considerations on sub-Hubble modes. It also differs from tachyonic preheating, i.e. where the instability is driven by the sign of ${\rm d}^2V/{\rm d}\phi^2$ dynamically turning negative, which was again found to be inconsequential for PBH production in models consistent with CMB data~\cite{Suyama:2006sr}. Finally, ``axion'' inflation coupled to gauge fields has also been revealed less prone to PBH production~\cite{Caravano:2022epk} than initially thought~\cite{Linde:2012bt,Garcia-Bellido:2016dkw}.}

This was presented as a possibility already in~\cite{Jedamzik:2010dq}, but as far as we can infer, it was not based on an analysis of the non-linear evolution of the modes. 
We remind the reader that a well-accepted criterion for BH formation is that the hoop conjecture should be verified~(see Ref.~\cite{Misner:1973prb}, pp.\ 867-868): the entirety of the object's mass must be compressed to the point that it can fit within a sphere whose radius is equal to that object's Schwarzschild radius. This is 
loosely equivalent to requiring that the Newtonian potential $|\Phi|\sim G{\cal M}/R$ associated to the mass ${\cal M}$ concentrated in a volume of radius $R$  attains $|\Phi|\sim 0.5$ (see e.g.~\cite{Lyth:2005ze}).\footnote{Some literature such as~\cite{Cotner:2018vug,del-Corral:2023apl} has sometimes estimated PBH formation {\it at sub-horizon scales} relying on different criteria, which do not appear generically reliable.} Notice that within a purely linear analysis, 
as the one in~\cite{Jedamzik:2010dq},
there is really no indication that the growth of density perturbations may lead to
BH formation, since the curvature perturbation $\mathcal{R}$ (reducing in modulus to the gravitational potential $\Phi$ on sub-Hubble scales, see footnote 2) stays constant: Put otherwise, extrapolating the linear behaviour into the non-linear regime (a procedure which is {\it not} anyway to be trusted~\cite{Parry:1998pn}) {\it is not indicative of} PBH formation. 

In the appendices of~\cite{Martin:2019nuw}, the allegedly rather generic formation of PBHs was supported by solving  the non-linear evolution equation of a spherically symmetric fluctuation of the energy density of the field, evolving in the homogeneous background of the scalar field itself, similarly to the approach followed in~\cite{Goncalves:2000nz}. As noted already in~\cite{Goncalves:2000nz,Martin:2019nuw}, the solved equations are rather idealized, notably assuming spherically symmetric configurations. In the case of dust, the pioneering studies on the subject~\cite{Khlopov:1980mg,Polnarev:1985btg,1985MNRAS.215..575K} showed that if and only if the fluctuations are homogeneous and spherically symmetric to a high-degree, the corresponding spherical symmetric solutions can be taken to represent reality.

A  number of articles have since been devoted to clarify the parametric 
dependence of the probability of the formation of (Hubble-sized) PBHs
on the root-mean-square relative fluctuation of density on a mass scale ${\cal M}$, $\sigma({\cal M})$, when asphericity and angular momentum are taken into account~\cite{Kuhnel:2016exn,Harada:2016mhb,Harada:2017fjm,Carr:2018nkm,Harada:2022xjp}.
In general, collapse is prevented by a number of effects: The isotropic approximation breaks down, gradients are important, non-Gaussian density and velocity effects are present, etc.

In fact, the analogous problem in cold dark matter cosmology is well-known,
and the dynamics (unless the power spectrum is extremely enhanced) leads to {\it halo formation} supported by velocity dispersion, rather than  to BHs.
The issue is then to understand to which extent something analogous happens also for the scalar field models of interest going beyond the dust-like fluid approximation, as noted e.g. in~\cite{Martin:2020fgl}.
An outcome qualitatively analogous to the cold dark matter case has actually been recovered in numerical simulations of early structure formation, neglecting GR effects, in a universe subject to a quadratic inflaton potential, if the post-inflationary metastable state is assumed to survive long enough~\cite{Musoke:2019ima}. No sign of large gravitational potentials, which would cast doubts on resorting to a Newtonian approximation, is found {there}.  
Not surprisingly, the corresponding halos have a profile resembling an NFW one~\cite{Eggemeier:2020zeg}.
Of course, since dealing with a bosonic degree of freedom, a solitonic condensate may form at the center of halos. Even in the rather idealised case where no self-interaction or other interactions 
are relevant,
for these objects to collapse into BHs, the  matter-dominated phase must last a very long time, equivalent to $\gtrsim 30\,$ e-folds~\cite{Eggemeier:2021smj}. Even neglecting the physical realism of these cases, definitely the PBH  outcome is far from generic.\footnote{Note that, as far as BH formation is concerned, the free bosonic case these simulations focus on is optimal, with self-interactions or interactions with other fields making formation of BH more challenging, see e.g.~\cite{Garani:2021gvc}.}

Is the GR treatment of gravity crucial to qualitatively alter these conclusions? Existing studies do not support this idea: For instance, the parametric family of potentials studied in~\cite{Kou:2019bbc} leads to BH formation only in a subset of cases when their parameter $\alpha<0.2$ (with $\alpha=1$ corresponding to the quadratic potential, see Fig.\ 8 in \cite{Kou:2019bbc}). The authors of~\cite{Muia:2019coe} also clearly state that the fate of non-linear structures forming in a quadratic inflaton potential is to disperse, not to end up forming BHs. This situation appears grim for PBH production, which are only found in special setups. In particular, we note that simulations leading to PBH formation start from already large density contrasts as initial conditions, whose dynamical plausibility in the post-inflationary era needs to be independently checked.
We will return to this point in Sec.~\ref{sec:lattice}.

\section{A more realistic scenario}
\label{sec:inflamodel}

In the previous section we worked with a purely quadratic monomial potential. This potential is ruled out\footnote{So is the $\phi^4$ potential  considered in~\cite{Jedamzik:2010dq}.} by Planck measurements of the CMB~\cite{Planck:2018vyg}, specifically due to the upper limit on tensor modes.  Yet, in previous literature~\cite{Jedamzik:2010dq,Martin:2019nuw,Auclair:2020csm}, it is implicitly or explicitly argued that this is a generic proxy for what happens in viable inflationary scenarios, since this quadratic potential is only meant to approximate the actual one in the post-inflationary era, when the inflaton oscillates around its minimum. However, in the self-resonant instability of interest here, the oscillating term in the Mathieu equation~\eqref{quadmathieu} comes from the tiny feedback from metric perturbations. In order for the self-resonance mechanism of refs.~\cite{Jedamzik:2010dq,Martin:2019nuw} to be considered generic, one has to show that these terms {\it systematically} dominate with respect to the terms due to the unavoidable anharmonic corrections involved in any phenomenologically viable single-field inflationary model. 
Considering two classes of potentials compatible with the CMB, we are going to show that this is not the case. More in general, our results suggest that metric preheating and a quadratic approximation for the potential cannot be simultaneously fulfilled.\footnote{{In this work we focus on single-field minimally coupled inflation. One could wonder if a purely quadratic potential might be a good approximation in other scenarios. If more scalars are involved in driving inflation, the dynamics of preheating is typically more complicated as well (so that a simple $\phi^2$ approximation is not sufficient), and the interactions between the fields tend to quench the (non-linear) growth of fluctuations. In scenarios like warm inflation, a quadratic potential is also excluded by the CMB for typical forms of the dissipation rate, see \cite{Ballesteros:2023dno}.}} 

\subsection{T-models}
\label{sec:Tmodel}

Let us start by focusing on a family of inflationary potentials called $\alpha$-attractor T-models~\cite{Kallosh_2013}, which has been motivated in the context of supergravity and has been extensively studied in the literature as they can provide a good fit to current CMB data, see e.g.\ \cite{Kallosh:2013yoa,Ferrara:2013rsa,Martin:2013nzq,Galante:2014ifa, CMB-S4:2016ple, Lozanov:2017hjm, Dalianis:2018frf}. This choice provides a concrete example that allows us to illustrate the main points of the physics we want to highlight. The inflaton potential of these models is parametrized as:
\begin{equation}
V(\phi) = \Lambda^{4} \text{tanh}^{2n} \left( \frac{|\phi |}{M} \right)\,,
\label{tmodel}
\end{equation}
where $n$, $\Lambda$ and $M$ are constants.
The scale $M$ separates a quasi-flat concave potential at  large field values from a post-inflationary monomial potential for small field field values. 
 For simplicity, let us take $n=1$ in this section.
If we match to a quadratic monomial potential in the small-field case, the potential reduces to
\begin{equation}
  V(\phi) \simeq 
    \begin{cases}
      \Lambda^{4} \left ( \frac{\phi}{M} \right )^{2} & \phi\ll M\\
      \Lambda^{4} & \phi\gg M
    \end{cases}       \,.
    \label{tmodellimit}
\end{equation}
The approximate plateau for $\phi\gg M$ is consistent with a power spectrum of primordial curvature fluctuations at CMB scales compatible with observations. In particular, the observational upper limit on the tensor-to-scalar ratio, $r<0.032$ at 95\% CL~\cite{Tristram:2021tvh}, requires $M\lesssim 5\,{\rm to}\, 10\, M_{\rm P}$, with the precise maximum allowed value of $M$ depending on the time elapsed between CMB scales leaving the horizon to the end of inflation.

The power spectrum of curvature perturbations $A_{s}$ at the fiducial scale $k_{*} = 0.05\, \text{Mpc}^{-1}$ is $\ln(10^{10}\,A_s)=3.044\pm0.014$ (from a Planck analysis of TT,TE,EE+lowE+lensing data)~\cite{Planck:2018vyg}. This sets the scale $\Lambda$: 
\begin{equation}
    \Lambda^{4}=\frac{3\pi^{2}A_{s}}{N_{*}^{2}}M^{2}\mp^{2}\,,
    \label{lambdaeq}
\end{equation}
where $N_*$ is the (e-fold) time at which the scale $k_*$ satisfies $k_* = a\,H$. In our analysis we adopt the value $N_{*}=55$, but our results do not depend qualitatively on this choice.
In the small-field limit, the effective mass squared is
\begin{equation}
m^2=2    \frac{\Lambda^{4}}{M^2}=\left(6.4\times 10^{-6}\frac{55}{N_*}\right)^2\mp^2\,\simeq (1.6 \times 10^{13}{\rm GeV})^2\,,
    \label{scalaron_m}
\end{equation}
hence the choice of the benchmark shown in Fig.~\ref{termsQuadPot}.

\begin{figure}[t]
\centering
\begin{tabular}{cc}
\includegraphics[width=0.75\textwidth]{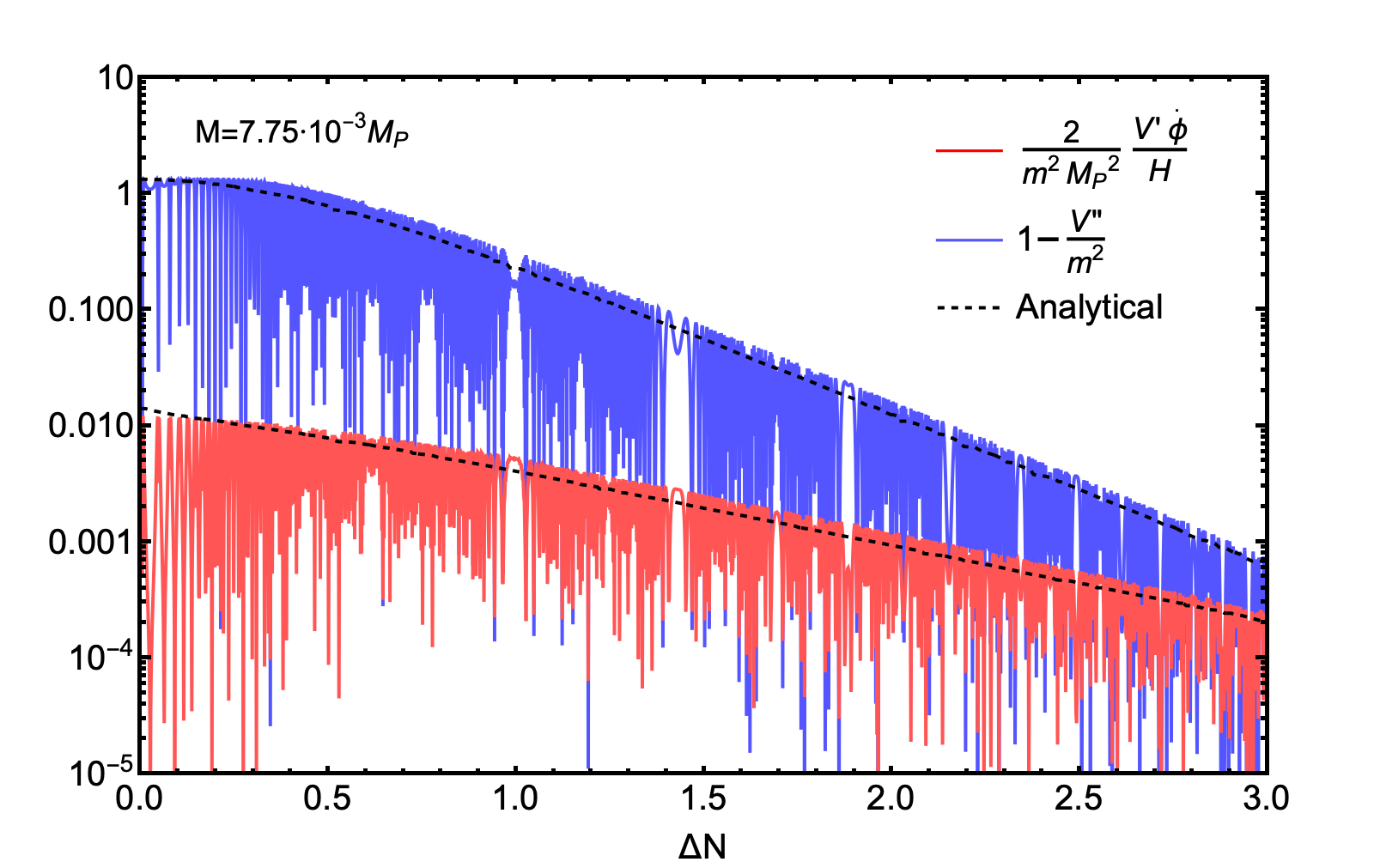}  
\end{tabular}
\caption{Comparison of the anharmonic term (blue) and the leading metric term (red) for the T-attractor potential assuming $M=7.75\times 10^{-3}\mp$ for the numerical solution. The analytical approximation corresponding to eq.~\eqref{as_solution} for each of the terms is indicated by a black dashed line, where we numerically solved for the scale factor.
}
\label{anharmT}
\end{figure}

Approximating the end of inflation by the condition $\epsilon_V= 1$, we get 
$\phi_{\rm end}\simeq \sqrt{2}\mp$ for $M\gtrsim \mp$, and $\phi_{\rm end}\simeq 0.5 M\ln(4\sqrt{2}\mp/M)$ for $M\ll \mp$, which suggests the initial post-inflationary range of interest for the field, $\phi_{\rm end}$, to be roughly given by the interval $[M,\mp]$. 
In Fig.~\ref{anharmT} we compare the anharmonic and the dominant metric terms (see eq.~\eq{Qfull1}): ${V''}/{m^2}-1$ and $2V' \dot{\phi}/(H\,m^2\mp^2)$
{during the first 3 post-inflationary e-folds, $\Delta N=0$ corresponding to $\epsilon=1$, with field values roughly spanning the range $\phi\in [0.024,1.13]M$, for the choice $M=7.75\times 10^{-3}\mp$ (the rationale for the choice $M\ll \mp$ will be soon clear). Here we denote with a prime the derivative with respect to $\phi$. The colored lines correspond to the numerical solution, while the dashed ones refer to the analytical approximation  valid at late times given by eq.~\eqref{as_solution}.}
The anharmonic term is much larger than the metric one, and therefore {\it metric} preheating is not occurring {for this choice of parameters. It is important to stress that the evolution depicted in Fig.~\ref{anharmT} cannot be meaningfully extrapolated to  many e-folds. Both the exponential amplification of the inflaton perturbations and the
subsequent backreaction (as we will see) take place within one e-fold after
inflation (see Fig.~\ref{oscplots} later on), at times where the assumption that the inflaton is undergoing damped oscillations in a quadratic potential is not justified.

\subsection{E-models}
Another popular class of inflationary potentials (still within the context of $\alpha$-attractors) is the the class of asymmetric $E-$models
of inflation~\cite{Kallosh:2013yoa}:
\begin{equation}
V(\phi) = \Lambda^{4} \left|1-\exp \left( -\frac{\phi }{M} \right)\right|^{2n}
\label{emodel}
\end{equation}
which, for $n=1$, has the same small-$\phi$ and large $\phi$ limits as in eq.~\eqref{tmodellimit} above, and coincides with Starobinsky's model of inflation~\cite{Starobinsky:1980te} for the choice $M=\sqrt{3/2} M_{\rm P}$.
Using $\epsilon_V=1$ to find the time of the approximate end of inflation, we obtain 
$\phi_{\rm end}\simeq \sqrt{2}\mp$ for $M\gtrsim \mp$ and $\phi_{\rm end}\simeq M\ln(\sqrt{2}\mp/M)$ for $M\ll \mp$, which again suggests the initial post-inflationary range of interest for the field, $\phi_{\rm end}$, to be roughly given by $[M,\mp]$; the same as for T-models. 
The effective mass squared in the small-field limit is again given by eq.~\eqref{scalaron_m}.
In Fig.~\ref{anharmE} we compare the anharmonic and the dominant metric terms, 
for $M=\mp$, which is complementary to the choice made for  
Fig.~\ref{anharmE}. 
We find now that the anharmonic terms are comparable, still marginally 
larger than the metric ones. 
{Metric-term dominance would require $M\gtrsim \mp$, which is only allowed in a small  parameter space while remaining phenomenological viable.
}

Furthermore, even in this case eq.~\eqref{as_solution} is valid only at late times while it is a poor approximation of the dynamics just after the end of inflation, where however the instability is already operational. We thus conclude that the quadratic approximation is nowhere reliable at a quantitative level.
The same argument applies also to the $\alpha$-attractor T-models.

\begin{figure}[htbp!]
\centering
\begin{tabular}{cc}
\includegraphics[width=0.75\textwidth]{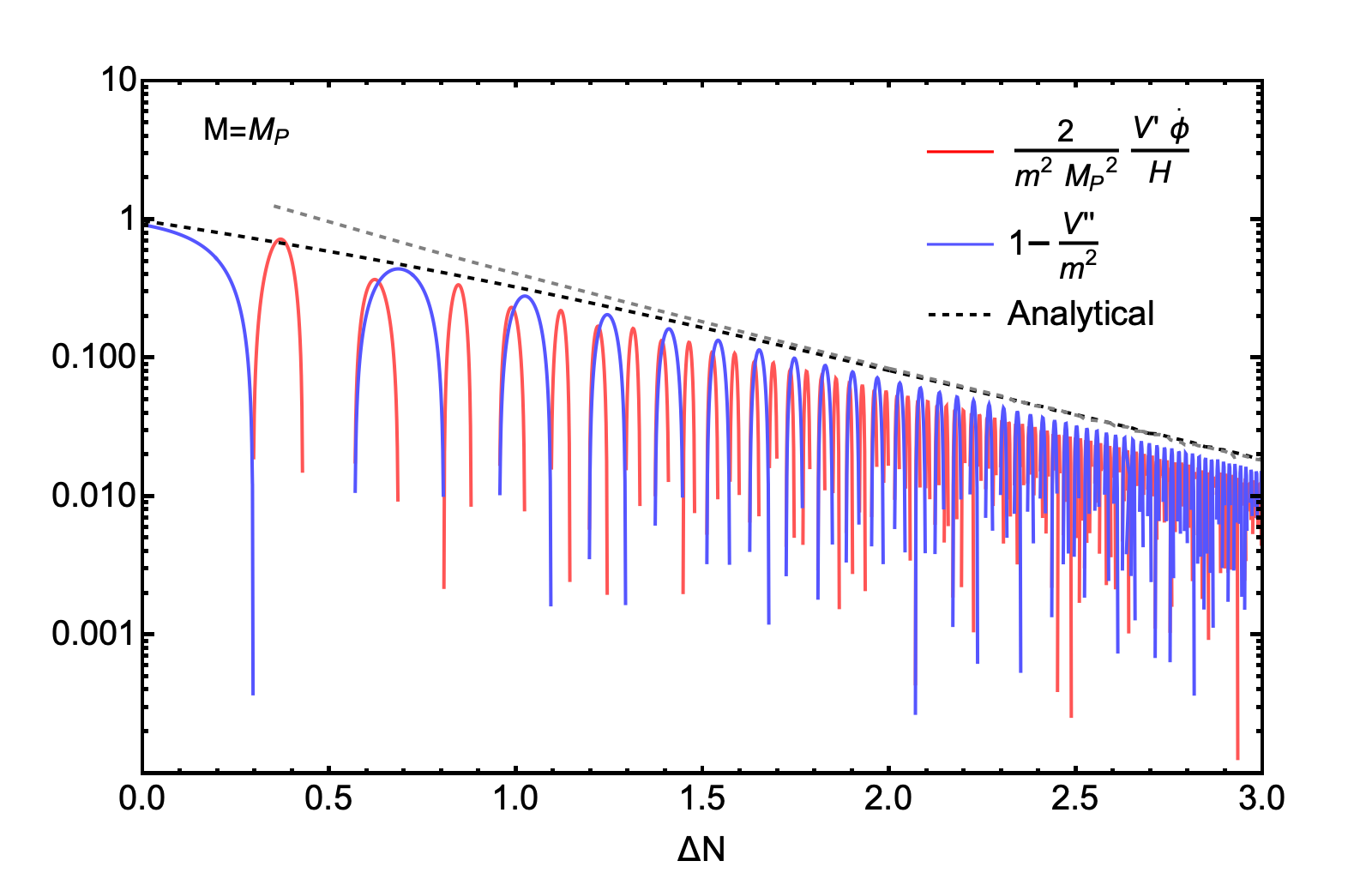}  
\end{tabular}
\caption{ Comparison of the anharmonic term (blue) and the leading metric term (red) for the E-attractor potential assuming $M=\mp$, close to the popular  Starobinsky model, for the numerical solution. The analytical approximation corresponding to eq.~\eqref{as_solution} is indicated by a black (anharmonic) and gray (metric) dashed line, where we numerically solved for the scale factor.}
\label{anharmE}
\end{figure}

\subsection{{The possible role of anharmonic terms}}

We have seen in the previous section that current knowledge suggests that a quadratic potential (recovered in our  potentials when $M\gg \mp$, being careful at the limited regime where it is viable) does not lead to PBH formation from metric-preheating. However, the results of this section may open an opportunity:  Since in most of the parameter space anharmonic terms are at least as important (when $M\sim \mp$), and often dominant (when $M<\mp$), with respect to metric ones, maybe one can rescue the idea of forming PBHs via a self-resonant instability, when anharmonicity is accounted for. 

A generalised form of Floquet's theory~\cite{Amin:2014eta} has been used in the past to identify which region of parameter space is most promising 
in enhancing the resonance, see e.g.~\cite{Amin:2011hj,Lozanov:2017hjm}. Rather intuitively, a fast and efficient non-linear growth corresponds to large values of the ratio $\Re(\mu_{k})m/H$. For the largest scales embracing the low-$k$ parameter space, one finds $\Re(\mu_{k})m/H\propto\mp/M$, with a coefficient depending on the potential shape~\cite{Amin:2011hj,Lozanov:2017hjm}. 
Then, in order to check if it is possible to rescue the possibility of forming PBHs by relying on an anharmonic self-resonance, one may want to look at the $M\ll \mp$ range, which we do next.

\section{Non-linear evolution: lattice results}
\label{sec:lattice}

Based on the previous considerations, the most promising case to consider for PBH formation from self-resonant preheating is the one characterised by  $M\ll\mp$, on which we will now focus.\footnote{
 Note  that the complementary, intermediate regime $0.01\lesssim M/M_P \lesssim 0.1$,which experience only moderate resonances, has been recently studied in~\cite{Eggemeier:2023nyu}. Notice that for sufficiently smaller values of $M$ tachyonic preheating may become more dominant than metric preheating, see~\cite{Tomberg:2021bll}. } In this case, a strong resonance regime ($\Re(\mu_{k})m/H\gtrsim 7$) is responsible for the amplification of the 
 perturbations. Furthermore, we have shown that in this case
 the dynamics leading to parametric amplification is dominantly driven by the anharmonic terms in the potential, rather than the metric perturbations.

In this section we {argue} that, while a fast growth of the fluctuations can be achieved, notably via the formation of metastable, non-perturbative {\it oscillons}, they stop growing well before PBH formation. Most of the results presented in this section are not original, but have been discussed in a different context than the one covered here, see e.g.~\cite{Lozanov:2017hjm,Lozanov:2019ylm}. Nonetheless, a survey of the PBH formation literature shows that they are often overlooked and somewhat unfamiliar, hence we find useful to re-discuss them here.

When nonlinear effects become important, the evolution of the perturbations can be dramatically modified with respect to linear theory results. A credible treatment of the relevant dynamics is crucial to determine whether gravitational collapse can occur and lead to the formation of PBHs. In order to study this regime, we employ the public software \texttt{CosmoLattice} (CL)~\cite{Figueroa_2021,Figueroa_2023}, which allows to perform lattice computations of the dynamics of interacting scalar and gauge fields in an expanding Universe.

\begin{figure}[t]
\centering
\includegraphics[width=1\textwidth]{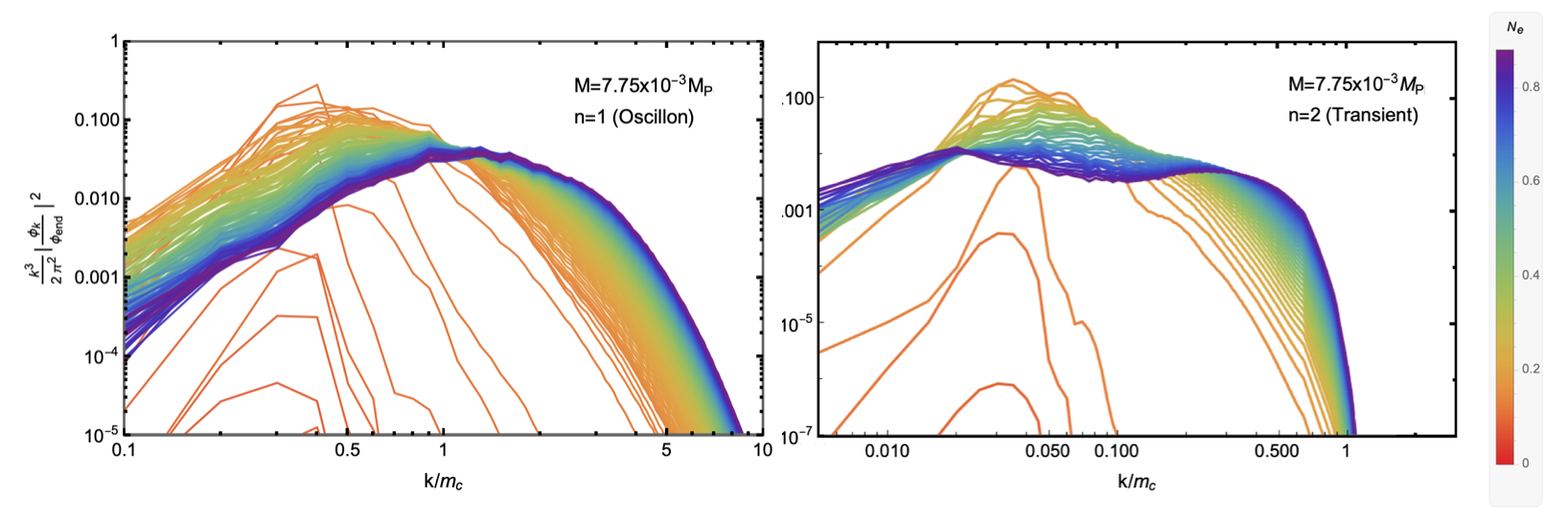}
\caption{Dimensionless power spectrum of the inflaton perturbations as a function of the comoving scale, see eq.~(\ref{eq:powerspectrum}). 
The color code denotes the time evolution (from red to violet) in the simulation.
Left ($n=1$) and right panels ($n=2$) are for two different values of the exponent in the scalar potential of the $\alpha$-attractor T-model in eq.~(\ref{tmodel}). These plots reproduce the results in reference~\cite{Lozanov:2017hjm}, in particular the ones in Figure 5 (left) and Figure 9 (left panel in the second row).
}
\label{oscplots}
\end{figure}

We follow the evolution of the inflaton field from the end of inflation and during approximately one e-fold.
The inflaton initial momentum can be estimated by the slow-roll dynamics before the end of inflation, although the results of the simulation are roughly independent on its exact value. 
Initial fluctuations of the inflaton field are set-up as explained in~\cite{Figueroa_2023}.
We present our results in Fig.~\ref{oscplots} for the T-models class of potentials, eq.~\eqref{tmodel}, a representative value of $M =7.75\times 10^{-3} \mp$, and for two cases: $n=1$ and $n=2$. We recall that the scale $\Lambda$ of the inflaton potential is fixed as in Eq.~(\ref{lambdaeq}), which is valid for any $n$. The figures show the evolution of the dimensionless power spectrum, defined as:
\begin{equation}
\mathcal{P}_{\phi}(k) = \frac{k^3}{2\pi^2} \left| \frac{\phi_{k}}{\phi_{\rm end}}\right|^{2},
\label{eq:powerspectrum}
\end{equation}
where $\phi_{k}$ is the Fourier mode of the inflaton field and $\phi_{\rm end}$ is the value of the inflaton field at the end of inflation,  approximated by the condition $\epsilon_{V}(\phi_{\rm end})=1$. 
We also define a generalised mass parameter
\begin{equation}
    m_{c}^{2} \equiv \frac{\partial_{\phi}V}{\phi}\bigg|_{\phi=\phi_{\rm end}} = 2n \Lambda^{2} \left(\frac{\Lambda}{M}\right)^{2}\left(\frac{\phi_{\rm end}}{M}\right)^{2n-2}\,,
\end{equation}
following the prescription in~\cite{Lozanov:2017hjm}, which reduces to eq.~\eqref{scalaron_m} for $n=1$. 

In the case $n=1$, fluctuations associated to long-wavelength comoving modes grow rapidly, leading to a broad peak in the primordial power spectrum. This behaviour corresponds to the parametric amplification of modes falling in the broad resonance band discussed in sec.~\ref{sec:resonance}. However, one can observe that the growth eventually stops in amplitude. This happens when the energy density stored in the perturbations becomes comparable to the one of the homogeneous condensate, and the backreaction on the dynamics becomes important.
At later times, the broad peak generated during the resonant phase just shifts towards larger values of the comoving scale, as visible in the left panel of Fig.~\ref{oscplots}. This is the manifestation of the formation of structures of fixed physical size, called {\it oscillons} in the literature, largely sustained by the {potential,}
see e.g.~\cite{Bogolyubsky:1976yu,Gleiser:1993pt,Copeland:1995fq,Honda:2001xg,Saffin:2006yk,Hindmarsh:2007jb,Amin:2010jq,Amin:2011hj,Amin:2013ika} for more details. The formation of these objects is related  to the oscillations of a scalar field in a potential which is quadratic around the origin and then flattens at larger field values, as the case $n=1$ in Eq.~(\ref{tmodellimit}). The results are qualitatively similar for $n=2$ (and in general $n>1$), see the right panel of Fig.~\ref{oscplots}, although in this case the broad peak does not shift as rigidly; the energy stored in these modes is thus redistributed, which corresponds to a decay of these structures, dubbed transients in~\cite{Lozanov:2017hjm}.
These results reproduce previous findings in the literature, see notably~\cite{Lozanov:2017hjm}.

Metric perturbations are neglected in these lattice simulations.  In the linear regime, this is justified for the parameter space of interest, as shown in sec.~\ref{sec:resonance}. As extensively argued in Ref.~\cite{Lozanov:2019ylm}, this is also the case at later stages of the evolution since the Newtonian potential $\Phi$ remains small. This can be understood making use of its relation to density fluctuations for modes that are deeply sub-Hubble: $\Phi\sim (a H/ k)^2 \delta\ll\delta$, where $\delta$ is the density perturbation. Importantly, since we are dealing with modes for which $k\gg a H$, this relation is gauge independent. So is the variable $\delta\phi$ describing the fluctuations of the inflaton field, which in this regime is just another way of expressing the Mukhanov-Sasaki variable $v$.
In particular, the gravitational potential on the surface of oscillons and transients is estimated to be $|\Phi| \lesssim 10^{-3},$
thus far from the values {$|\Phi| \sim \mathcal{O}(1)$}
required to form a BH.
Hence, for $M \ll M_{\rm P}$ the parametric growth of the inflaton perturbations during preheating is important, and, as we already discussed in Sec.~\ref{sec:inflamodel}, it corresponds to the regime where anharmonic terms dominate over metric contributions, justifying resorting to CL, where the latter are neglected. In this regime, parametric instability leads to the formation of weakly gravitating metastable objects, which are not expected to collapse into PBHs, rather to eventually dissipate. This suggests that the production of PBHs from preheating is far from generic in phenomenologically viable, well-motivated, and simple inflationary scenarios, such as the one we have considered. 

PBH formation is at least in principle possible when $M\sim M_{\rm P}$, as suggested by numerical relativity simulations in~\cite{Muia:2019coe,Nazari:2020fmk,Kou:2019bbc} {(see also~\cite{Padilla:2021zgm,Hidalgo:2022yed} for related analytical work)}.  However, the question of whether the needed conditions for PBH formation are generically found in self-consistent inflationary models is rarely addressed. Instead, parametrically set initial conditions for the inflaton overdensity are typically adopted.  For instance, quoting from Ref.~\cite{Muia:2019coe}: 

{\it \ldots the objects we consider in this work have a compactness comparable to that of the corresponding black hole. The formation of such objects needs to be checked for each specific model via dedicated lattice simulations. In the simplest and most model independent scenarios, self-interactions of a single field are sufficient to make the quantum fluctuations grow and enter the non-linear regime.} 

In Fig.~\ref{in_cond_02Mp} we do perform such a test with a lattice calculation, similar to the case shown in Fig.~\ref{oscplots}, but adopting $M=0.2\mp$ to mimic the parameter choice adopted in~\cite{Muia:2019coe}. Note that for this choice, the metric backreaction is still sub-leading, and the lattice results should still be at least qualitatively reliable. Our results show that the enhancement of the power-spectrum, while present, is $\sim 8$ orders of magnitude weaker compared to what would be needed for the simulations of ref.~\cite{Muia:2019coe} leading to PBH formation to be considered very representative. Indeed $\mathcal{O}(1)$ values of the metric potential are required for collapse into a black hole.

Recent work performed for plateau-like inflationary models in full GR simulations~\cite{Aurrekoetxea:2023jwd} agrees with these conclusions: The compactness of the object formed (a proxy for $|\Phi|$) reaches a maximum value of $\sim 0.01$ when the equivalent of our parameter $M$ attains $M\sim 0.05\,{\rm to}\, 0.1\,M_{\rm P}$. No BH forms if inhomogeneities are comparable to typical inflationary fluctuations seeds.

At the current state of knowledge, all the indications confirm  our expectation that, in agreement with intuition, very specific conditions on the inflaton potential are actually needed to trigger PBH formation in the post-inflationary era: The regime where parametric self-resonance is prominent is in tension with the requirement of forming PBHs, at least in single field models. 

\begin{figure}[t]
\centering
\includegraphics[width=0.75\textwidth]{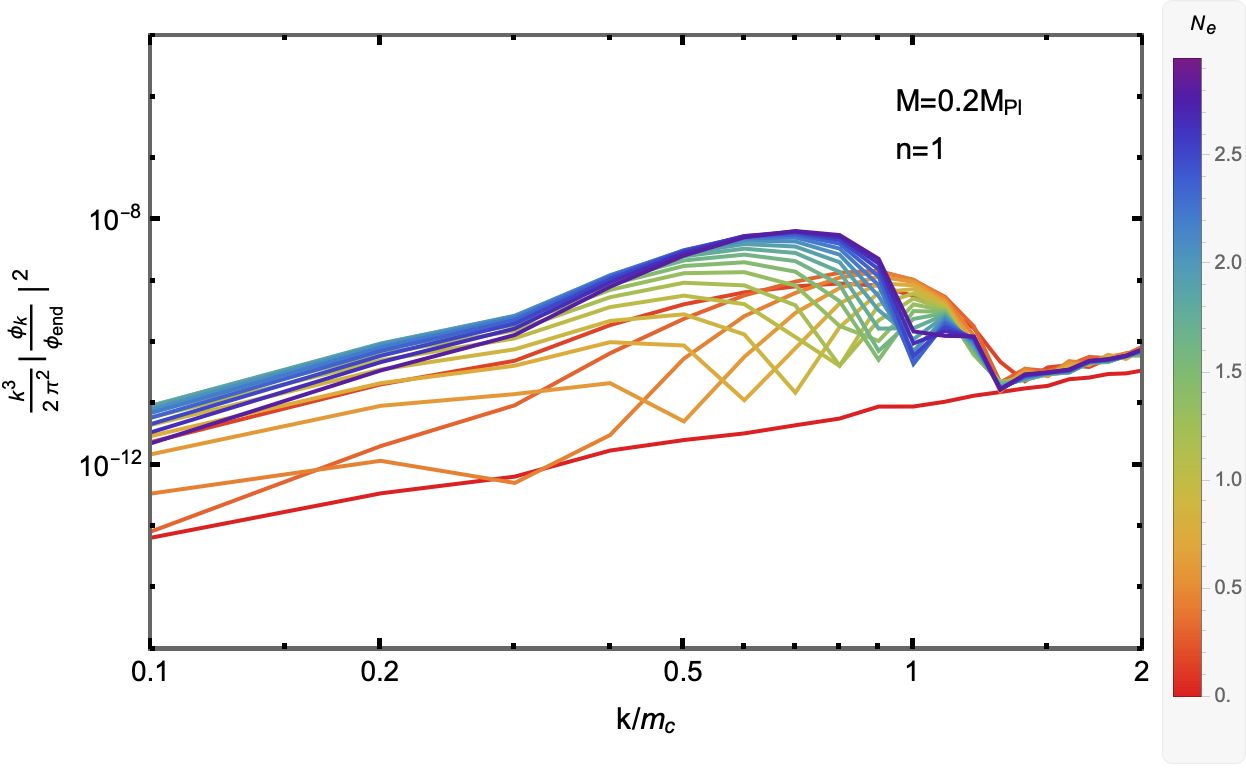}
\caption{Lattice calculation, similar to the case shown in Fig.~\ref{oscplots}, but adopting $M=0.2\mp$.}
\label{in_cond_02Mp}
\end{figure}

\section{Conclusions}
\label{sec:conclusions}

Despite their theoretical and phenomenological interest, the production of PBHs is rather challenging from the model-building point of view. Interestingly, it has been  claimed that sizeable PBH formation arises quite generically during preheating, thanks to the self-resonant phenomenon triggered by metric terms~\cite{Jedamzik:2010dq,Martin:2019nuw,Auclair:2020csm}. In this era, the inflaton field oscillates around the bottom of its potential with a decreasing amplitude. 
In these articles, it was assumed that
a quadratic approximation of the potential is generically sufficient to describe the preheating dynamics.
At the background level, the Universe behaves approximately as in a matter-dominated era, since the effective (averaged over fast oscillations) equation of state of the inflation field is close to the one of a pressureless fluid, $w\simeq0.$ Crucially, perturbations of sub-Hubble modes are amplified by a parametric resonant effect sourced by the oscillation of the inflaton field itself. This happens for modes inside the corresponding instability band, see eq.~\eqref{sub-hor-inst}. 
We have reanalysed this important issue, recently also discussed in more extended models~\cite{del-Corral:2023apl}, and found significantly less promising results.

As a preliminary remark,  it is worth stressing that the fluctuation enhancement one can obtain is at best\footnote{This has to do with two facts: i) the average equation of state is slightly larger than zero, so that the growth of fluctuations is slightly less efficient than in matter domination, since $\delta \propto a^{1-3w}$ and $\langle w \rangle>0$ (see Appendix~\ref{formulae});
ii) generalised Floquet theory shows that only some wavelength bands, among all the ones allowed by the Jeans criterion, are actually allowed to grow, see the discussion around 
eqs.~\eqref{sub-hor-inst},\eqref{Jeans}.} equivalent to the standard Newtonian growth of perturbations in a matter-dominated expansion phase. 
A scrutiny of the literature shows that forming PBHs in a matter-driven expansion phase is much more challenging and definitely not generic. We attribute this discrepancy with the original publications on the topic to a too simplistic treatment of the non-linear growth of perturbations, that neglects {non-linear} feedback (and have been treated in a too idealised setting); in some other articles, much more optimistic conclusions can be traced back to an unjustified application of horizon-scale (energy density contrast) criteria for collapse into PBHs to sub-horizon fluctuations.

Furthermore, we have argued that a treatment of metric preheating independent of the specifics of inflaton potential, i.e.\ limited to the quadratic approximation, can not be {\it generically} used for reliable quantitative studies. In fact, in realistic models anharmonic terms in the potential can be sizeable and dominate the dynamics describing the evolution of the density perturbations. For the sake of concreteness, we have demonstrated this result in the context of the $\alpha$-attractor T-models and E-models. In general, we could not identify a situation where the quadratic approximation and the metric preheating dominance are simultaneously verified to quantitatively rely on the approximate treatment common in the literature.

Based on this insight, we have investigated with lattice calculations if the anharmonicity can be used to rescue the original idea, i.e.\ to resort to inflaton self-interactions to boost the formation of non-linear structures, thus easing PBH formation. {\it We find that this is not the case}, since the parameter space where the formation of non-linear structures is significantly enhanced (corresponding to the quadratic to flat potential transition scale $M$ being significantly smaller than $\mp$, and where GR effects are unimportant) corresponds to the one where these objects are metastable oscillons, rather than PBHs. We have recovered results presented in the context of early universe oscillon physics and rarely appreciated in the context of PBH literature.  

Based on a few GR numerical studies~\cite{Muia:2019coe,Nazari:2020fmk,Kou:2019bbc,deJong:2021bbo}, the least challenging parameter space to form PBHs by self-resonance in the post-inflationary phase seems to correspond to the range $M\lesssim \mp$.  We have raised awareness on the fact that even in this case, it may be difficult to obtain the needed large inhomogeneous initial conditions from the mere quantum fluctuations induced from generic and phenomenologically viable inflationary potentials subject to parametric amplification, confirming the results recently presented in~\cite{Aurrekoetxea:2023jwd}. 
A further exploration of self-consistent scenarios fulfilling these conditions certainly deserves further study, but it is already clear that PBH formation is not generic even in this case. 

Despite the largely negative nature of the results we obtained, we would like to stress that the  post-inflationary scalar field dynamics remains a rich and interesting topic, which has been explored and still deserves further exploration concerning a number of phenomena, such as  primordial gravitational waves,  magnetic fields, non-gaussianities, baryogenesis, and dark matter, beyond the topic of PBH production~\cite{Amin:2014eta}. Even limiting oneself to PBHs, besides further numerical GR studies, 
another possibility to form sub-horizon scale PBH relies on the interaction of oscillons/solitons  both  via gravity and new interactions, as reviewed in~\cite{Flores:2024eyy}. Existing studies on gravitational scattering of oscillons/solitons after preheating show some interesting dynamical clustering, but do not provide any  indication of PBH formation, suggesting that if PBH formation happens, it must be a rare process~\cite{Amin:2019ums}.
We would then argue that our conclusion that PBH formation is not generic should be read as an opportunity, in case of PBH detection, to shed light on very peculiar dynamics in the early Universe. 

\section*{Acknowledgments}
The work of GB has been funded by the following grants: 1) Contrato de Atracci\'on de Talento (Modalidad 1) de la Comunidad de Madrid (Spain), 2017-T1/TIC-5520 and 2021-5A/TIC-20957, 2)  PID2021-124704NB-I00 funded by MCIN/AEI/10.13039/501100011033 and by ERDF A way of making Europe, 3) CNS2022-135613 MICIU/AEI/10.13039/501100011033 and by the European Union NextGenerationEU/PRTR, 4) the IFT Centro de Excelencia Severo
Ochoa Grant CEX2020-001007-S, funded by MCIN/AEI/10.13039/501100011033.
MT acknowledges support from the research grant “Addressing systematic uncertainties in searches for dark matter No.
2022F2843’ funded by MIUR and the project ``Theoretical Astroparticle Physics (TAsP)'' funded by INFN. We would like to thank Mustafa A.\ Amin, Marcos A.\ G.\ Garc\'ia, and Vincent Vennin for feedback on a preliminary version of this manuscript.

\noindent

\appendix

\section{Post-inflationary matter-like expansion: A closer inspection}
\label{formulae}

Depending on the applications, the approximation of using the perfect fluid, dust-like equation of state for the scalar field density and pressure may prove inadequate, and the actual dynamics may deserve going beyond that approximation. See for instance~\cite{Martin:2020fgl} for an example of the changes associated to the presence of the radiation background the inflaton starts decaying into. Here, we analyse the equation of state and the sound speed associated to the homogeneous inflaton field oscillating around the minimum of a quadratic potential after inflation.
Considering the asymptotic behaviour in eq.~\eqref{as_solution}, the energy density $\rho(t)$ and pressure density $P(t)$ are:

\begin{equation}\label{eqrhoan}
\rho(t)=\bar{\rho}(t)\left\{1+\frac{9}{4}\frac{H^2}{m^2}\cos^2[m\,(t-t_{\rm end})]+\frac{3}{2} \frac{H}{m}\sin[2m\,(t-t_{\rm end})]\right\} \,,
\end{equation}
and
\begin{equation}\label{eqPan}
P(t)= \bar{\rho}(t)\left\{-\cos[2m\,(t-t_{\rm end})]+\frac{9}{4}\frac{H^2}{m^2}\cos^2[m\,(t-t_{\rm end})]+\frac{3}{2} \frac{H}{m}\sin[2m\,(t-t_{\rm end})]\right\} \,,
\end{equation}
where
\begin{equation}
\bar{\rho}(t)\equiv m^2\frac{\phi_{\rm end}^2}{2}\left(\frac{a_{\rm end}}{a}\right)^{3}\,.
\label{rho_bar}
\end{equation}

The standard approximation consists in assuming $\rho\simeq \bar{\rho},\, P\simeq 0$, or its generalisation to non-quadratic potentials. 
More in general, averaging over a time ${\cal T}$, one can define an average equation of state parameter $\langle w\rangle_{\cal T}\equiv \langle P\rangle_{\cal T}/\langle \rho\rangle_{\cal T}$~\cite{Cembranos:2015oya}. 
If we define the oscillation period $T=2\pi/m$, since in the regime of our interest $m \gg H$, one can consider $a(t)$ and $H(t)$ constant during time intervals equal or longer than the oscillation time. Under this approximation one gets:
\begin{equation}
\langle w\rangle_T=\frac{\langle P\rangle_T}{\langle \rho\rangle_T}\simeq\frac{9}{8}\frac{H^2}{m^2}\label{avw}\,.
\end{equation}

Similarly, one can compute the average of the square of the sound speed $\langle c_s^2\rangle_T=\langle \dot P\rangle_T/\langle \dot \rho\rangle_T$ (see e.g. sec. II in~\cite{Bean:2003fb} and ref.~\cite{Hertzberg:2014iza}) obtaining:

\begin{equation}\label{avcs2}
\langle c_s^2\rangle_T=\frac{\langle \dot P\rangle_T}{\langle \dot \rho\rangle_T}
\simeq \frac{9}{4}\frac{H^2}{m^2}=2w\,.
\end{equation}
It is worth noting that the average quantities eqs.~(\ref{avw},~\ref{avcs2}), agree
with the prediction
\begin{equation}
c_s^2=w-\frac{\dot w}{3(1+w)H}\,,
\end{equation}
which can be derived from the Friedmann equations.

In practice, these simple analytical averages do not capture the leading corrections to the approximation $\rho=\bar\rho$ and $P=0$. 
In fact, one should take into account that i) the scale factor evolves approximately as $a(t)\sim t^{2/3}$, ii) the solution in eq.~\eqref{as_solution} is approximate.
Both effects are of the order ${\cal O}(H/m)$, i.e. larger than the correction obtained above~\cite{Cembranos:2015oya}.

\begin{figure}[t]
\centering
\includegraphics[width=0.75\textwidth]{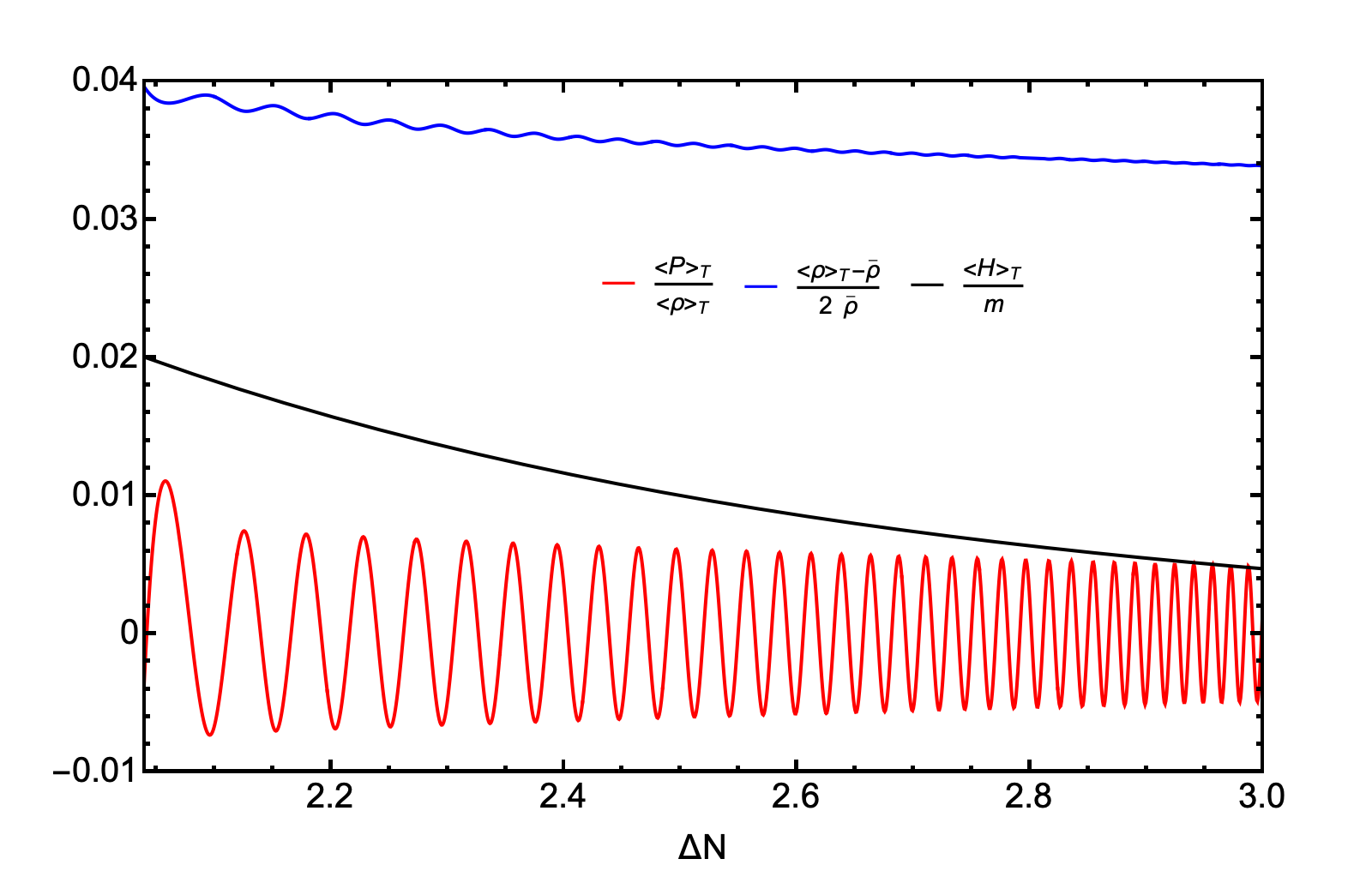}
\caption{The numerically derived $\langle w\rangle_T$ (red), (half of the) relative difference of the numerical average energy density and eq.~\eqref{rho_bar} (blue), and the average of the Hubble rate over the inflaton mass $m$ (black). All averages have been computed over a single inflaton oscillation period $T=2 \pi /m$, and are shown over the third e-fold in the post-inflationary era, in order to illustrate a ``late time'' behaviour.}
\label{averages_app}
\end{figure}

Since this correction is rarely illustrated in the literature, in  fig.~\ref{averages_app} we gauge the departure of the time-average of $\rho$ with respect to $\bar\rho(t)$ (blue curve). Considering that for this example $H/m\sim{\cal O}(0.01)$, we see that the actual corrections  to the obtained averaged $\rho(t)$ are of order $H/m$ (black curve), with both effects (i), (ii) mentioned above contributing to the same level.
Despite the highly oscillatory behaviour of the pressure, its (asymmetric) oscillations and thus $\langle w\rangle_T$ (red curve) are also of order $H/m$, rather than $H^2/m^2$ as from eq.~\eqref{avw}.

\bibliographystyle{utphys}
\bibliography{biblio}  

\end{document}